\documentclass{appolb}
\usepackage{graphicx}
\usepackage{mhchem}
\usepackage{amsmath}
\usepackage{amssymb}
\usepackage{enumitem}
\usepackage{setspace}  
\usepackage[T1]{fontenc}
\usepackage[english]{babel}
\usepackage[utf8]{inputenc}
\usepackage{bbm}
\usepackage{bm}
\usepackage{mhchem}
\usepackage{microtype}

\usepackage{hyperref}
\usepackage[
backend=biber,
style=phys,
biblabel=brackets,
citestyle=numeric-comp,
sorting=none,
minnames=4,
maxnames=4,
]{biblatex}
\DeclareLanguageMapping{polish}{english}

\numberwithin{equation}{section}

\AtBeginDocument{%
\renewcommand{\figurename}{Fig.}
}

\begin{document}


\title{A nonstandard statistics for strongly correlated systems: Two simple examples
\thanks{Presented at Conference {\it Concepts in Strongly Correlated Quantum Matter Conference} (CSCQM), Kraków, Poland, 20–22 November, 2025.}}
\author{
Józef Spałek, Piotr Kuterba,  
\address{Department of Condensed Matter Theory and Nanophysics, \\ Institute of Theoretical Physics, \\ Jagiellonian University, 30-348 Kraków, ul. Łojasiewicza 11 \\}
Mateusz Wójcik, \\
\address{AGH University of Kraków, 30-059 Kraków, al. Adama Mickiewicza 30\\}
Danuta Goc-Jagło,
Maciej Fidrysiak,\\ 
\address{Department of Condensed Matter Theory and Nanophysics, \\ Institute of Theoretical Physics, \\ Jagiellonian University, 30-348 Kraków, ul. Łojasiewicza 11 \\}
Leszek Spałek,\\ 
\address{Department of Materials \& Metallurgy, University of Cambridge, 27 \\ Charles Babbage 
Road,   Cambridge CB3 0FS, UK\\}
Włodzimierz Wójcik\\
\address{Institute of Physics,
Faculty of Materials Engineering and Physics,\\ Cracow University of Technology, 31-155 Kraków, ul. Podchorążych 1}
}
\maketitle
\newpage
\begin{abstract}
We discuss two different cases of strongly correlated fermions statistics. The first of them is the non-Fermi liquid (NFL) case, $i.e.$, that of fermions with exclusion of doubly occupancy of quasimomentum states $\{\bf k\}$ with opposite spins ($\uparrow\downarrow$). The second is that of statistical spin liquid (SSL) case, in which the fermion spins hop around and mix with the holes (unoccupied) states. For both cases we calculate the system entropy and the corresponding statistical distribution function, analyzed for a two-dimensional square-lattice filling $n\in [0,1]$ and relative temperature $k_B T/|t|$, where $t<0$ is the nearest neighbor hopping integral. We are particularly interested in the situation when the system of itinerant fermions reduce to the Mott-insulator state for the half-filled band ($n\to1$). This
limiting situation signals a qualitative difference between the present SSL statistics and that of uncorrelated fermions representing normal Fermi-liquid state.
\end{abstract}
\section{Introduction}
The quantum statistical physics of bosons \cite{Spalek2020a} and fermions \cite{Fermi1926} was established almost
simultaneously with the birth of the quantum mechanics (1925-1926). 
Its formal approach relies on the method introduced originally by Boltzmann and Gibbs of optimizing the number W of configurations of $N_e$ particles among micro $N_e$-particle states, in accordance with macro conditions, that
their total energy E and total number $N_e$ are conserved. The crucial is also the fact that the number of configurations for particles with energies $\{ \epsilon_{\bf k} \}$ for given complete set of possible values of quantum numbers  $\{{\bf k}\}$ is highly degenerate (i.e., with the degeneracy $g_{\bf k} \gg 1$); this particular circumstance simplifies essentially the calculation procedure using the Stirling formula for expressing the factorials of particle numbers $\{N_{\bf k}\}$ and empty $\{ N - N_{\bf k}\}$ states, as well as treating the variables $\{N_{\bf k}\}$ as a quasicontinuous set. The optimal number of configurations (distributions) is  then expressed solely by the explicit knowledge of the energy spectrum of individual particles $\{\epsilon_{\bf k}\}$ and by the two principles: Pauli exclusion principle for fermions, as well as their indistinguishability in both cases. Both of these rules have its common origin \cite{Dirac1926} by employing the particle {\it indistinguishability principle, explicitly} introduced by Dirac symmetry (for bosons) and antisymmetry (for fermions) of $N_e$-particle wave function with respect to transposition of the pairs of particles, i.e., with respect to their coordinates and spin or the complete sets of their quantum numbers interchange, ${\bf k} \longleftrightarrow {\bf k}'$. In result, the optimal (most probable) particle distribution function $n_{\bf k} \equiv N_{\bf k}/g_{\bf k}$ is obtained for the corresponding ideal quantum gases in the form 
\begin{align}\label{eq:1}
n_{\bf k} \equiv \frac{N_{\bf k}}{g_{\bf k}} = \frac{1}{ \exp\left[\beta(\epsilon_{\bf k} - \mu)\right] + \eta},
\end{align}
where $\beta \equiv (k_B T)^{-1}$ is inverse absolute temperature T (in energy units), $k_B$ is the Boltzmann constant, $\mu$ is the so-called chemical potential obtained from the condition of total particle conservation, $\sum_{\bf k} N_{\bf k} = N_e$, and $\eta = \pm 1$ for fermions and bosons, respectively.  
From the explicit knowledge of (\ref{eq:1}), all the physical properties such
as the functions of state and related quantities can be determined.

 The basic question we have posed  some time ago \cite{Spalek1988d,Spalek1990d} (and thoroughly reexamined in our present group) is whether such a program, with assuming only the general form of $\{\epsilon_{\bf k}\}$, can be generalized to the situation of strongly correlated particles, e.g., by introducing additional constraints due to the fact that now interaction not only renormalizes their bare starting energies $\epsilon_{\bf k}$, but first of all, removes some of the states (double occupancies) from the Fock space of $N_e$ particles. We think that posing such a question and
 solving it may be helpful also for a detailed discussion of concrete experimental results for those systems. At least, it may contribute to a platform for discussing their universal (semi)quantitative features.


Explicitly, we construct our scheme based on the Hubbard lattice models with double occupancy  exclusion of the individual states $\{{\bf k}\}$ in the reciprocal space of particles with the opposite spins \cite{Byczuk1994,Spalek1988d1,Hatsugai1992,Byczuk1995,Vittoriano2005}. This type of principle roughly 
represents the situation encountered in the Hubbard model with infinite-range repulsive interaction \cite{Byczuk1995,Vittoriano2005} and, in this manner represents a sort of complementary situation to the original Hubbard model \cite{Hubbard1963} formulated in real space, which is based on the intraatomic limit of repulsive interaction between lattice spin-$1/2$ fermions.
However, in this paper we consider here only on the basic aspects solely from the point of view of statistical physics, leaving the microscopic quantum and statistical properties  to a separate discussion.

\section{Two examples of nonstandard statistics}

In this section we analyze two specific examples of introducing nonstandard statistics for strongly correlated fermions. Both of them are regarded as belonging to the non-Fermi (non-Landau) separate classes of quantum liquids strongly correlated fermions with double-occupancy exclusion of quasimomentum states of particles with opposite spins.

\subsection{Non-Fermi liquid (NFL)}
As we said earlier, we exclude the doubly occupied quasimomentum states $|{\bf k}\rangle$ for particles with energy $\epsilon_{\bf k}$,
here considered on a square lattice of fermions with the particle energy taken in the tight-binding approximation

\begin{equation}
    \epsilon_{\bf k} = 2 t (cos k_x + cos k_y) + 4t'\cos k_x\cos k_y,
\end{equation}
where $t<0$ and $t'>0$ are the amplitudes of the nearest-neighbor and next-nearest-neighbor hoppings, respectively. In what
follows we take $t=-1$ as the unit of energy and discuss the properties for both $t'/|t| = 0.25$ and variable of  band filling $n\equiv \sum_{\mathbf{k}} N_{\bf k}/N$ and temperature T.
Note that the first Brillouin-zone in dimensionless quasimomentum space is defined by $-\pi \leq k_x , k_y \leq \pi $, and the corresponding limits of the band
filling in this single-band in the uncorrelated case varies in the range $0\leq n\leq 2$. However, as here we exclude completely 
the double occupancies, it actually can vary only in the interval $n \in [0,1]$.

In accordance with our exclusion of $\{ | {\bf k } \uparrow \downarrow \rangle \}$ states in addition to the 
usual Pauli principle, the number of macro configuration W is 
\begin{equation}
    W = \prod_{{\bf k} \sigma} \binom{ g_{\bf k} - N_{{\bf k }  \downarrow}}{N_{{\bf k }  \uparrow}} \binom{ g_{\bf k} - N_{{\bf k }  \uparrow}}{N_{{\bf k }  \downarrow}}. 
\end{equation}
The non-Fermi liquid character of the system is connected with the presence of the extra term $N_{{\bf k} \overline{\sigma}}$ in the respective numerators of the Newton symbol. 

Next, we define functional $\mathcal{F} \{ N_{{\bf k} \sigma} \}$ to be minimized with respect to $\{ N_{{\bf k} \sigma} \}$, i.e. 
\begin{equation}
    \mathcal{F} \equiv k_B \ln W - \alpha_0 \left(\sum_{{\bf k} \sigma} N_{{\bf k} \sigma} - N \right) - \alpha_1 \left(\sum_{{\bf k} \sigma} \epsilon_{\bf k} N_{{\bf k} \sigma} - E \right).
\end{equation}
Employing the standard procedure \cite{Spalek1988d,Spalek1990d,Byczuk1995} we obtain the following expression for 
the system entropy in this case
\begin{align}
    S \equiv S_{NFL} & = k_B \sum_{{\bf k} \sigma} g_{\bf k} \left[(1- n_{{\bf k} {\sigma}}) \ln (1- n_{{\bf k} {\sigma}}) - n_{{\bf k} {\sigma}} \ln n_{{\bf k} {\sigma}} \nonumber \right. \\  & \left. -  (1- n_{{\bf k} }) \ln (1- n_{{\bf k} })\right].
\end{align}

Note that this expression differs remarkably from the corresponding one for the noninteracting fermions.
For the particular case of paramagnetic state $n_{{\bf k} \sigma} = n_{{\bf k} \overline{\sigma}}$ the statistical distribution function takes the explicit form
\begin{equation}\label{eq:6}
    n_{\bf k} = 1 - \frac{1}{\sqrt{1 + 4 \cdot \exp(- \beta (\epsilon_{\bf k} - \bar{\mu}))}}.
\end{equation}
with effective chemical potential $\tilde{\mu} \equiv \mu + k_B T \ln 2 $. The additional term in $\bar{\mu}$
increases its value beyond the band limits, in addition to its formal form, distinguishing it from the Fermi-Dirac
(FD) distribution.

Detailed discussion of this statistics is carried out in the next section, after an analytic elaboration of the statistical spin liquid (SSL) concept below.

\subsection{Statistical spin liquid (SSL)}
Some time ago two of the authors \cite{Spalek1988d} introduced the concept of statistical spin liquid (SSL) composed of hopping spins on the  lattice of fermionic particles mixed up with the holes in a partially filled ($n \leq 1$)
narrow band of fermionic nature. In a simplest manner, the number of macro configurations for such a liquid would take the form
\begin{equation}\label{eq:7}
    W = \prod_{\bf k} \binom{g_{\bf k}}{N_{\bf k}} 2^{N_{\bf k}},
\end{equation}
where the number of holes is $N_h = N - N_e$. One can see immediately that entropy such a liquid in the limit of Mott insulator ($n=1$) reduces to 
\begin{equation}
    S = k_B \ln W = N k_B \ln 2.
\end{equation}
Above we have assumed that for $n=1$, $g_{\bf k} = N_{\bf k}$: This limiting value differs from the corresponding expression for ordinary fermion (FD case),
which is then $S = 2 N k_B \ln 2$, as in the latter situation double ${\bf k}$ occupancies of particles with opposite spins are not excluded.

In the expression (\ref{eq:7}) the number of fermionic configuration $\binom{g_{\bf k}}{N_{\bf k}}$ is counted as independent of that due to the spin degrees of freedom, $2^{N_e}$. This means that the expression represents
state with charge-spin separation, as happens precisely in correlated one-dimensional systems \cite{Lieb1968}.
However, to start with a proper fermion picture of correlated particles we have postulated \cite{Spalek1988d,Spalek1990d} that, in general, the expression the following form replacing (\ref{eq:7}) should be
\begin{equation}\label{eq:9}
    W = \prod_{\bf k} \binom{g_{\bf k}}{N_{\bf k}} \frac{N_k !}{N_{{\bf k} \uparrow}!  N_{{\bf k} \downarrow}!}.
\end{equation}
This expression represents a two-fluid system with spin $\sigma = \uparrow$ and $\downarrow$, intermixed with the holes. Again, we assume that the components with $\sigma = \uparrow$ and $\downarrow$ are distinguishable in the macroscopic 
state when, e.g., the system is polarized magnetically. Such a situation appears certainly when we have a strongly correlated system with the spin-direction dependent effective masses \cite{Spalek1990a,Spalek2006a} (treated in detail in a separate work).  Here, as in (\ref{eq:7}) the
spin subsystems $\uparrow$ and $\downarrow$ are distinguishable in general situation.

The postulated form (\ref{eq:9}) leads in a straightforward manner to the entropy expression
\begin{equation}
    S \equiv S_{SSL} = - {k_B} \sum_{ {\bf k}} \left[ (1 - n_{\bf k}) \ln(1 - n_{\bf k}) + n_{{\bf k} \uparrow} \ln n_{{\bf k} \uparrow} + n_{{\bf k} \downarrow} \ln n_{{\bf k} \downarrow}\right]
\end{equation}
and to the particle statistical function
\begin{equation}
    {n}_{{\bf k} \sigma} = (1 - {n}_{{ {\bf k}} \overline{\sigma}}) \frac{1}{1 + \exp[\beta(\epsilon_{{\bf k} \sigma} - \mu)]} 
\end{equation}
The entropy expression reduces to the value $S = N k_B \ln 2$ in the Mott insulating limit.
In the next section we compare the relevant properties in the two cases (a) and (b)) and their relation to the  statistical properties of ordinary (FD) fermions.
\begin{figure}[!htbp] 
\centering
\includegraphics[width=0.7\textwidth]{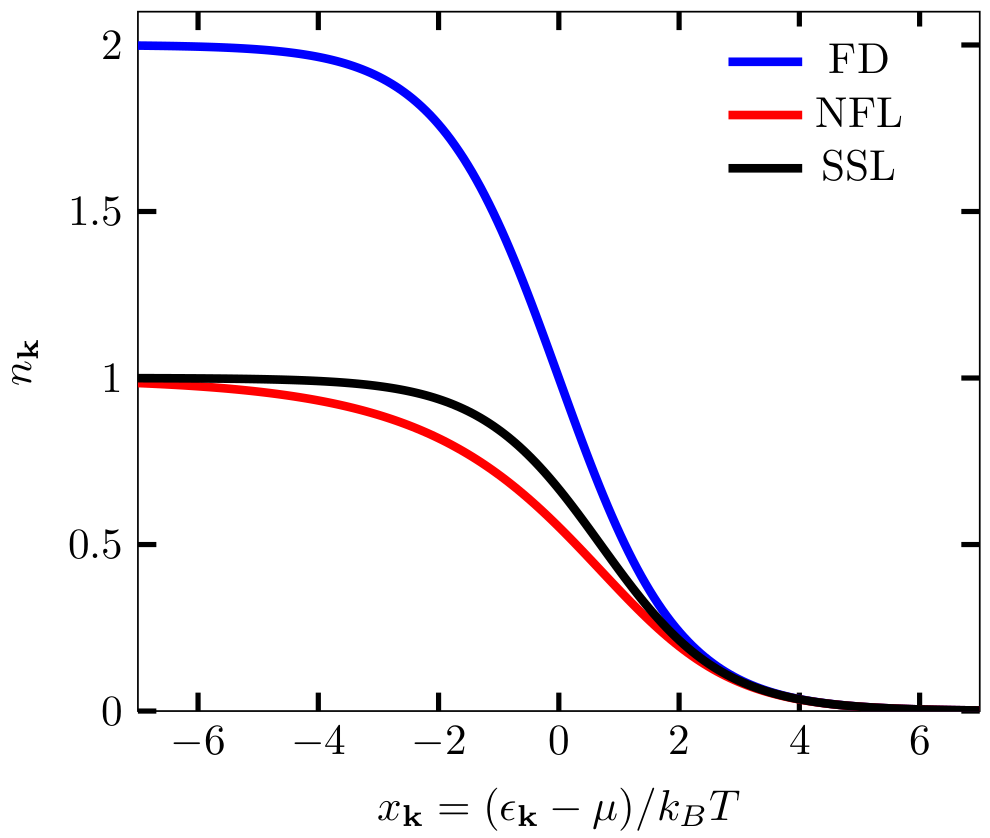}
\caption{Overall behavior of the statistical distribution function $n_{\bf k}$ as a function of relative particle energy $x_{\bf k} \equiv (\epsilon_{\bf k} - \mu)/k_B T$ for three
cases: Fermi-Dirac (FD, blue line), non-Fermi liquid (NFL, red), and statistical spin liquid 
(SSL). For detailed explanation see main text.}\label{fig:1}
\end{figure}

\section{Properties and comparison of NFL and SSL cases}

\begin{figure}[!htbp] 
\centering
\includegraphics[width=0.8\textwidth]{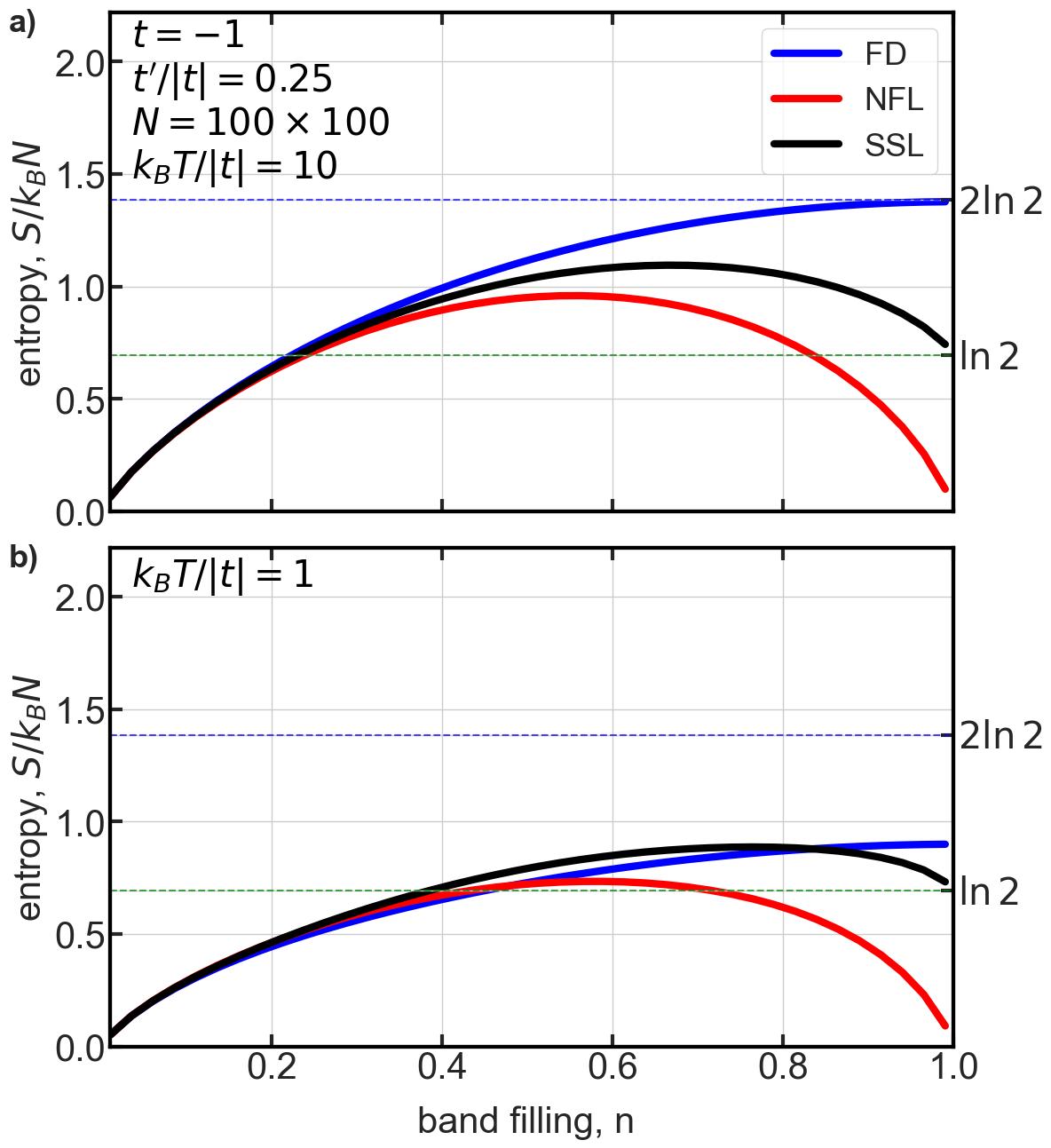}
\caption{Entropy {\it versus} band filling n for the three statistics: FD, NFL, and SSL
for the two temperatures $k_B T/|t|$: $10$ and $1$. Note that for $n\to 1$, the entropy reaches the local-moment limit $\ln 2$ for SSL, whereas the corresponding high-temperature limit in the FD cases is $2 \ln 2$, and for NFL $S\to 0$. }\label{fig:2}
\end{figure}
In Fig. \ref{fig:1} we draw statistical distribution functions as a function of relative energy $(\epsilon_{\bf k} - \mu)/k_B \equiv x_{\bf k}$ for the three cases: Fermi-Dirac ideal gas (FD), NFL, and SSL. The overall behavior on this scale is similar, except that in FD total occupancy far below $\mu$ reaches the maximal value of two, whereas in the remaining two situations it is equal to unity. This circumstance illustrates directly that the double occupancy exclusion in the latter two cases. Note also that for $x_{\bf k} > 1$ the functions behave in a similar manner and represent the highly diluted gas, for which the additional exclusion is only marginally relevant. 

\begin{figure}[!htbp] 
\centering
\includegraphics[width=0.7\textwidth]{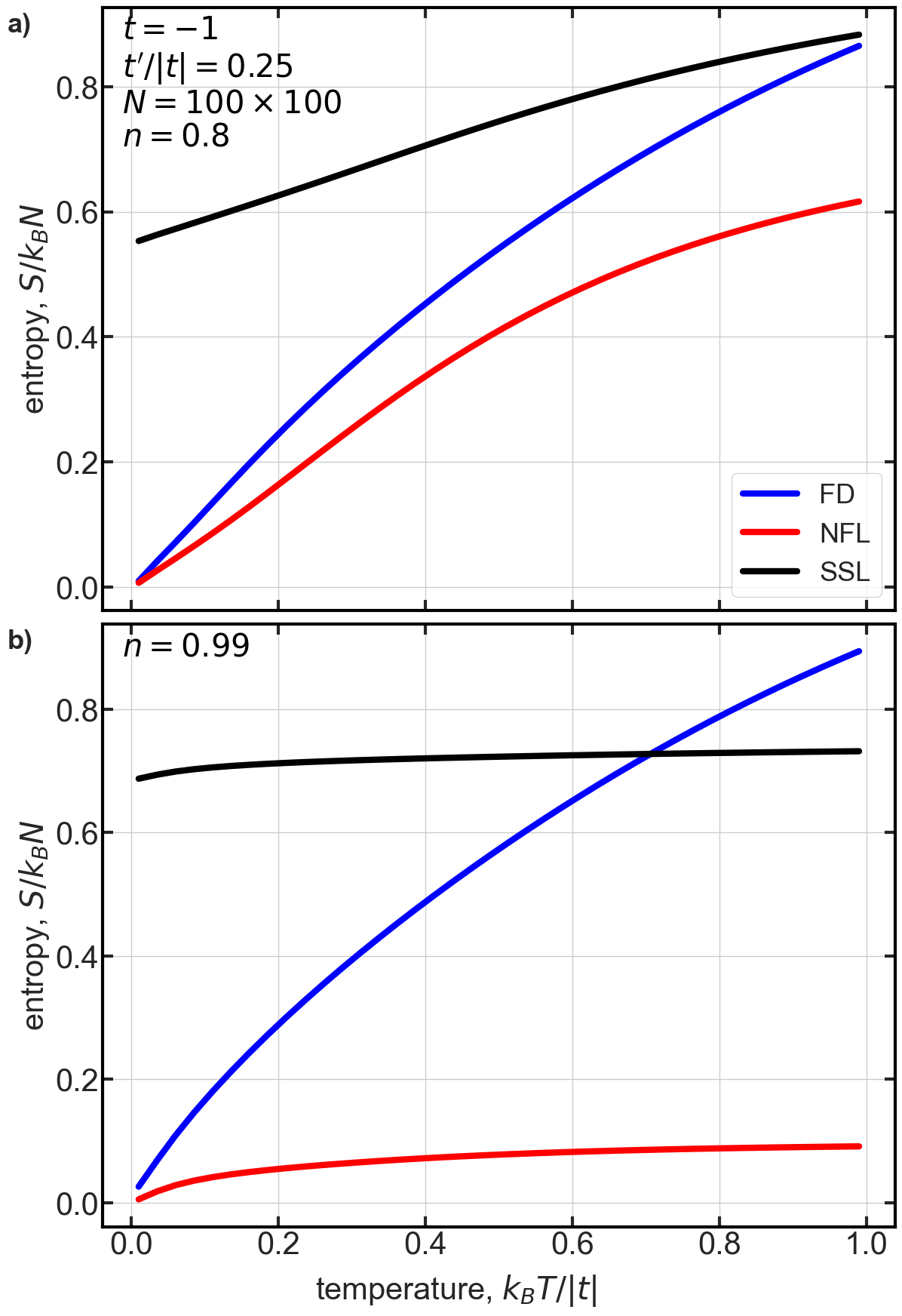}
\caption{ Entropy {\it versus} temperature for the two band fillings, $n=0.80 ~ ({\rm a})$ and $0.99 ~({\rm b})$. 
For $n\to1$ the full local moment is recovered only in SSL case, a clear sign of the statistics changeover from that of itinerant (indistinguishable particles) to that of local character (distinguishable spins) in ${\bf k}$-space.}
\label{fig:3}
\end{figure}

To illuminate the difference between the NFL and SSL we plot in Fig. \ref{fig:2} the entropy (in dimensionless units and per site) for the three cases as a function of filling $n = \sum_{{\bf k} \sigma} N_{{\bf k} \sigma} / N$, for two different temperatures specified, both in the high-temperature regime. As said above, only SSL case provides a proper value $\ln 2$ in the limit of Mott insulator ($n=1$). Obviously, for $n\to 0 $ the behavior is the same in all cases, as the additional exclusion does not play any role in the limit of diluted gas. However, there is a difference in their behavior in the low-T limit, as shown in Fig. \ref{fig:3} on example of two fillings $n=0.8$ and $n=0.99$. Even though in 
both cases the system is not far from the half-filling, in the SSL case, the entropy approaches its maximal value $\ln 2$. Also, the entropy $S_{SSL} \neq 0$ even when $T\to 0$, which means that the mixture of the two spin liquids is incoherent in the quantum-mechanical sense (note the presence of the Boltzmann factor in (\ref{eq:9}), reflecting the classical 
mixing of those two  liquids). This particle distinguishability with respect to the spin orientation will be discussed separately. 

\begin{figure}[!htbp]
\centering
\includegraphics[width=0.7\textwidth]{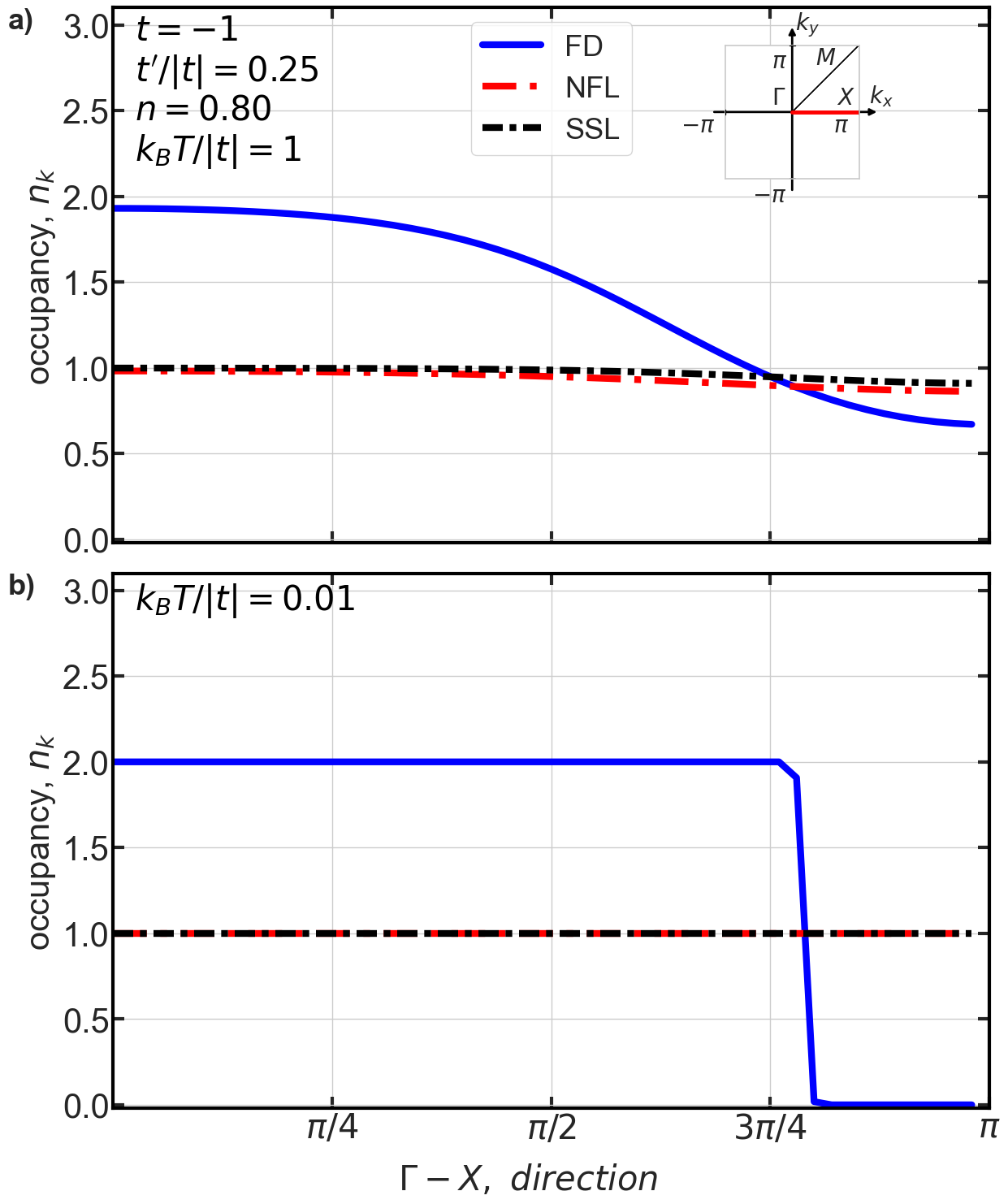}
\caption{Statistical distribution function $n_{\bf k}$ {\it versus} quasimomentum ${\bf k} = (k_x, 0)$; i.e., along  $\Gamma-X$ direction in the Brillouin zone (cf. inset). Temperature $k_BT/|t| = 1 $ (a) and $0.01$ (b). Note that chemical potential position in the NFL and SSL cases touches the zone boundary ($k_x, \pi$), whereas it is placed at about $k_x=\pi/2$, as elaborated
in the text.} \label{fig:4}
\end{figure}

\begin{figure}[!htbp] 
\centering
\includegraphics[width=0.7\textwidth]{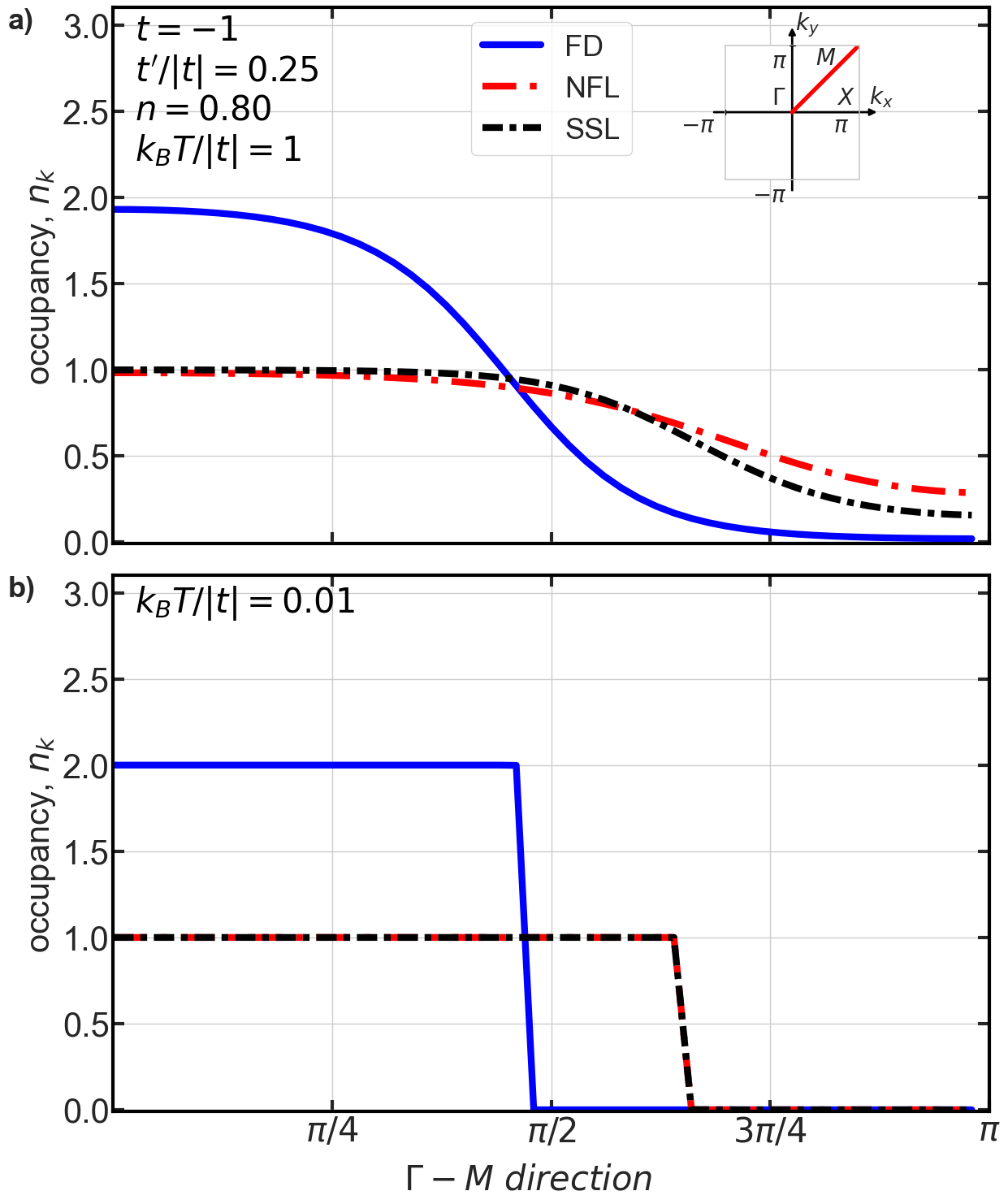}
\caption{Distribution functions $n_{\bf k}$ along the $k_x = k_y ~ (\Gamma-M)$ direction, for two
temperatures specified. The difference between the chemical potential position for NFL and SSL
from one side and that for FD is clearly visible, particularly at lower temperature. For
elaboration of this difference see main text.}\label{fig:5}
\end{figure}

Figs. \ref{fig:4} and \ref{fig:5} provide a detailed statistical distributions $n_{\bf k}$ in 
high- and low-T regimes, $k_B T/|t| = $ 1 and 0.01, respectively. Furthermore, the contours in ${\bf k}-$space taken,  along which the distributions, in ${\bf k} - $space were traced, are those with $k_x \neq 0$, $k_y = 0$
($\Gamma-X$ direction, cf. Fig. \ref{fig:4}, in the Brillouin zone) and $k_x = k_y$ ($\Gamma - M$ direction, cf. Fig. \ref{fig:5}),  respectively. Those directions are shown explicitly in the inset and characterize the Brillouin 
zone for the square lattice of fermions. Note also that the Fermi level is higher in SSL and NFL  cases and this circumstance is due to the exclusion part of the states (those with double occupancies) in the NFL and SSL cases. Also,
in this respect, it is not easy to distinguish between the two distributions, particularly near their Fermi-level position. In effect, close enough to the Mott insulator limit their distinction is not well resolved. Obviously, in the limit of Mott insulator the particles become distinguishable strictly as their classical location is defined, i.e., that on the atomic sites.
\begin{figure}[!htbp] 
\centering
\includegraphics[width=0.6\textwidth]{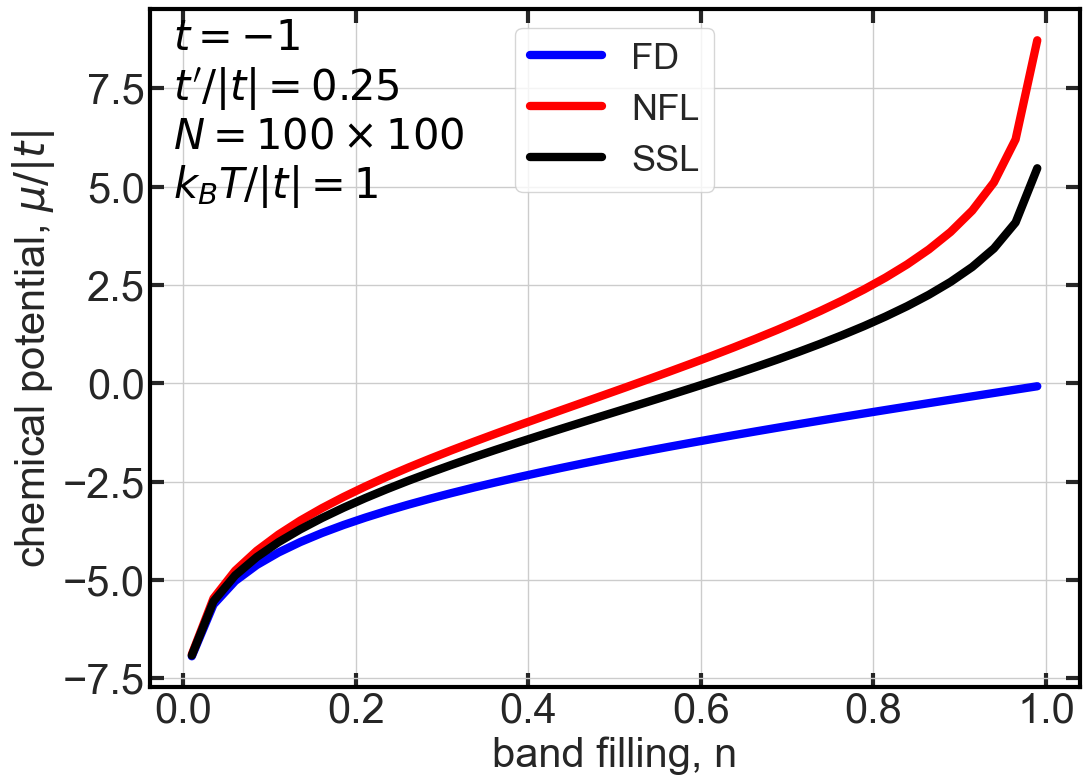}
\caption{ Typical chemical potential evolution {\it versus} band filling. Note the unusual behavior 
in NFL and SSL cases, as it exceeds the band lower and upper edges. In the latter two cases $\mu$ cannot be interpreted as the Fermi energy in the same sense as in the FD case.}\label{fig:6}
\end{figure}

Finally, in Fig. \ref{fig:6} we plot the position of the chemical potential in the three cases, all as the function of the band filling. In the FD case $\mu \to 0$ as $n\to1$, i.e., it is located in the middle point of the band, because in the FD case each ${\bf k}$ state can accommodate up to two fermions (with opposite spins)). On the contrary,
the position of $\mu$ increases spectacularly as $n\to1$ in SSL and NFL cases and reflects a strong restriction on the
particle occupancy close to the Mott insulating boundary.

\section{Conclusions}
From all this detailed analysis one feature should be underlined. Namely, the fact that the statistics of
correlated itinerant fermions in the SSL case changes into that of localized moments in the Mott insulator
limit, a feature characteristic for the systems undergoing the metal-insulator transition of the Mott-type
\cite{Spalek1987}. This transmutation of statistics, particularly when combined with the question of transition from indistinguishable to distinguishable particles accompanying it, should have principal consequences to our understanding of the strongly correlated systems as a new universality class of quantum matter, which encompasses both the liquid and localized nature of the carriers composing it, into a single picture.

\section*{Acknowledgments}
The research was support by Narodowe Centrum Nauki (NCN) Grant Nos. UMO-2021/41/B/ST3/04070 and 2023/49/B/ST3/03545.
\section*{Data Availability}
The dataset containing simulation results presented in this work is available in
Ref. \cite{Spaleketal2026data}.
\clearpage
\printbibliography
\end{document}